\documentclass[aps,amsmath,amssymb,pre,notitlepage,twocolumn,showpacs]{revtex4}
\usepackage{float}
\usepackage{color}
\usepackage{graphicx}

\renewcommand{\vec}[1]{\mathbf{#1}}

\begin{document}
\title{{Average balance equations, scale dependence, and energy cascade for granular materials}}
\author{Riccardo Artoni}
\email{riccardo.artoni@ifsttar.fr}              
\affiliation{LUNAM Universit\'e, IFSTTAR, MAST/GPEM, Route de Bouaye CS4 44344 Bouguenais France.}

\author{Patrick Richard}\email{patrick.richard@ifsttar.fr}          
\affiliation{LUNAM Universit\'e, IFSTTAR, MAST/GPEM, Route de Bouaye CS4 44344 Bouguenais France.}

\date{\today}

\begin{abstract}
A new averaging method linking discrete to continuum variables of granular materials is developed and used to derive average balance equations. 
Its novelty lies in the choice of the decomposition between mean values and fluctuations of properties which takes into account the effect of gradients. Thanks to a local homogeneity hypothesis, whose validity is discussed, simplified balance equations are obtained. 
This original approach solves the problem of dependence of some variables on the size of the averaging domain obtained in previous approaches which can lead to huge relative errors (several hundred percentages). It also clearly separates affine and nonaffine fields in the balance equations. The resulting energy cascade picture is discussed, with a particular focus on unidirectional steady and fully developed flows for which it appears that the contact terms are dissipated locally unlike the kinetic terms which contribute to a nonlocal balance.  
Application of the method is demonstrated
in the determination of the macroscopic properties such as volume fraction, velocity, stress, and energy 
 of a simple shear flow, where the discrete results are generated by means of discrete particle simulation.

\end{abstract} 


\maketitle
\section{Introduction}

The behavior of granular materials under deformation and flow is often studied by means of numerical methods that solve the equations of motion for every single body and account for the interactions between particles through appropriate models (see, e.g., Refs.~\cite{Silbert2001,Richard_PRL_2008,Rycroft2009}).
These methods, commonly grouped under the name of discrete element methods (DEM)~\cite{Frenkel2001,Schafer1996,luding2008}, can be extremely useful  for simulating discrete media with the purpose of establishing micro-macro relations, and in this perspective they need consistent averaging procedures~\cite{babic97,Glasser_PoF_2001,zhu02,Weinhart_PoF_2013,Weinhart_granularmatter_2012b,Weinhart_granular_matter_2012}.
An averaging method computes continuum-like variables (e.g., a velocity field) starting from their discrete, particle-scale counterparts (e.g., particle velocities). When using such an approach, two issues have to be considered with utmost attention: the representativity of the average and the physical meaning of the obtained estimates.
We can say that an average performed at a given point is representative if a sufficiently large domain in space and time can be defined, where particles share the same properties. Such a volume, usually called ``representative volume element," may be defined with respect to only some variables.

An important step in the development of averaging techniques was made by Babic~\cite{babic97}. In that work the author developed a general framework for weighted space-time averages, applicable to a wide range of conditions, and giving a large set of self-consistent continuum balance equations. In  Babic's method, the definition of a representative volume is intrinsically that of a zone where affine velocity fields are locally uniform, i.e., no gradients occur.
The assumption that such a volume can  be defined is called by Babic the ``continuum assumption." Now, the lack of scale separation typical of granular flows (the scale of spatial variation of variables has the same order of magnitude of the particle size) implies that in a three-dimensional (3D) flow we cannot in practice define such a volume. As a consequence, the physical meaning of averages obtained using Babic's approach in a 3D flow is generally questionable. At least some average estimates will then probably depend on the size of the averaging volume~\cite{Glasser_PoF_2001,Xu2004}.

We can see a simple example of this by considering a uniform shear flow; 
if all particles follow the mean flow with no frustration, we will tend to say 
that fluctuations are negligible. But this depends on the definition of fluctuation.
 If we calculate the fluctuational kinetic energy $\varepsilon^T$ [which is related to granular temperature~\cite{savage84}; see also Eq.~(\ref{eqn:epsilonT})]
  by considering fluctuations 
 as Babic does, we find  $\varepsilon^T\sim (\dot \gamma D)^2$, where $D$ is the averaging scale.  Thus the fluctuations depend on the averaging scale which casts doubt on such a method.
Also the kinetic contribution to the stress tensor will suffer the same problem. Thus, as we will show below, such an approach may lead to errors that can reach several hundred percentages.

In this work we focus on this problem, with a particular attention on the mechanical energy equation. We  discuss both
 the issue of dependence on the size of the averaging domain and
  the energy cascade picture associated with the averaging method.

It is clear that the dependence of some average variables on the averaging domain size is intrinsic to Babic's definition of 
fluctuations. In the following we show how a different fluctuation decomposition taking into account 
gradients yields a new derivation of continuum balance equations.
These balance equations have the merit of clearly separating affine and nonaffine fields 
and identifying the terms responsible for averaging scale dependence.
With this decomposition it is possible to attribute a more sound physical meaning to the 
terms appearing in the final balance equations.
We show how these balance equations may be greatly simplified by a ``local homogeneity assumption,"
 which is strongly related to the fluctuation decomposition itself and which yields less strict requirements than Babic's ``continuum assumption."
Applying this method to a simple class of flows, we find that two separate paths for fluctuating energy dissipation are implied: 
while kinetic stress power enters a nonlocal  balance, the contact stress power seems to be dissipated locally. 
This result, as well as the robustness of the method, is verified with discrete element simulations of a simple shear flow.

\section{A new derivation of balance equations}
\subsection{Weighted average, fluctuation decomposition and local homogeneity assumption}
Given a particle property $\psi_p (\vec x_p,t')$, Babic defined the averaged property $\bar \psi$ 
at point $\vec x$ and time $t$ as a weighted space-time average:
\begin{equation}
\rho \bar \psi = \int_{-\infty}^{\infty}\sum_p w(\vec x_p -\vec x, t'-t) m_p  \psi_p dt',
\end{equation}
where $w(\vec x_p -\vec x, t'-t)=w_p$ is a normalized weighting function, $\rho$ the average density, and $m_p$ the mass of a grain.
The fluctuation was simply defined by Babic with respect to the center of the averaging volume:
\begin{equation}
\tilde \psi_p= \psi_p -\bar \psi (\vec x).
\end{equation}

This may seem harmless, but in fact--- as shown in Ref.~\cite{Glasser_PoF_2001}---the dependence of some terms of the continuum averaged equations
 on the size of the averaging domain is strictly related to this choice for the fluctuation decomposition. 

We propose to define the fluctuation not with respect to the average value at point  $\vec x$ (as was done in Ref.~\cite{babic97}), 
but with respect to the particle center,
\begin{equation}
\tilde \psi_p= \psi_p -\bar \psi (\vec x_p).
\end{equation}
In the following we will redevelop continuum-averaged equations in the spirit of Babic's approach,
 but using this new fluctuation decomposition.

In order to simplify the result, we will make a constitutive assumption.
 We suppose that a scale exists where the gradients of  $ \bar \psi $ are smooth
  and that if this scale is the averaging scale intrinsic to the weighting function $w_p$, 
  we can approximate $\nabla  \bar \psi $ as a constant near the averaging point. 
  We call this assumption the \textit{local homogeneity assumption}. Together
   with this assumption we also assume a  homogeneous distribution of particle centers near the averaging point 
and the decorrelation of positions and velocities. These three assumptions are strictly related.\\

While the absence of scale separation usually prevents identification of a scale where gradients are zero, 
 such a first-order approximation is reasonable and---as we will see---very fruitful.

The local homogeneity assumption, involving gradients instead of variables, 
is less strict than Babic's continuum hypothesis and therefore more likely to be valid under 
a proper choice of the weighting function. 
Based on this assumption, we can therefore
approximate $ \bar \psi (\vec x_p)$ by
\begin{equation}
\bar \psi (\vec x_p)= \bar \psi (\vec x) + (\vec x_p-\vec x) \cdot \nabla \bar \psi (\vec x).
\end{equation}

Following Babic's approach, it is possible to develop a generic balance equation for the particle property $ \bar \psi (\vec x_p)$. Details are given in  Appendix~A; here we show the main findings, with a particular
regard to the continuity, momentum, and translational energy equations.
When performing the derivation, many new terms appear (with respect to Babic's treatment),
but most of them vanish due to the local homogeneity hypothesis.

\subsection{Continuum balances}

The local homogeneity hypothesis leaves the continuity equation  unchanged:
\begin{equation}\label{conti}
\frac{\partial}{\partial t} \rho +\nabla \cdot \rho \bar{\vec{ v}}  =0,
\end{equation}

where average density and velocity are defined by:

\begin{equation}
\rho= \int_{-\infty}^{\infty} \sum_p w_p m_p  dt',
\end{equation}

\begin{equation}
\rho \bar{\vec{ v}}= \int_{-\infty}^{\infty} \sum_p w_p m_p \vec{v}_p dt'.
\end{equation}

As for the momentum balance, we end up with the classical equation:
\begin{equation}\label{momentum}
 \frac{\partial }{\partial t} \rho \bar {\vec{ v}}+ \nabla \cdot \rho \bar{\vec{ v}}
   \bar{\vec{ v}}=\nabla \cdot \mathbf T +\rho \vec g,
\end{equation}
where the total stress tensor is composed by three contributions: $ \mathbf T =  \mathbf T^c + \mathbf T^k + \mathbf T^\gamma $. 
The first two terms are the same as in Babic. The first contribution is due to contact forces (both long-lasting contacts and collisions) and is given by 
\begin{equation}\label{tc}
\mathbf{T}^c =  \int_{-\infty}^{\infty}\sum_p\sum_{q>p}  w^F_{pq}  \vec{l}_{pq} \vec f_{pq}dt',
\end{equation}
where $w^F_{pq}$ is a weight function related to the fraction of the branch vector joining the two particles $p$ and $q$ lying within the averaging function.
Due to the new decomposition, Babic's kinetic stress tensor is decomposed in a truly kinetic part,
\begin{equation}\label{tk}
\mathbf{T}^k = - \int_{-\infty}^{\infty}\sum_p  w_p m_p \vec{\tilde {v}}_p \vec{\tilde {v}}_p dt' ,
\end{equation}
and a new term,
\begin{equation}\label{tgamma}
\mathbf{T}^\gamma = - \rho (\vec D \cdot \nabla  \vec{\bar{v}})(\vec D \cdot \nabla \vec{\bar{v}}),
\end{equation}
which contains all the dependence on averaging domain size through the vector $\vec D$. {It is useful to recall that the emergence of a part of the stress tensor depending on velocity gradients has nothing to do with constitutive relations but is the joint product of coarse graining and scale dependence. Due to their definition, both $\mathbf T^\gamma$ and $\mathbf T^k$ are symmetric. Evidently, in absence of velocity gradients, $T^\gamma$ will vanish; in that case, the kinetic stress tensor as defined by Babic will no longer depend on the coarse-graining length. The components of $\vec{D}$,}
 related to the characteristic 
size of the domain along each direction, 
are defined by the following equation:

\begin{equation}
\rho D_i^2  = \int_{-\infty}^{\infty} \sum_p w_p m_p  (x_{pi}-x_i)^2 dt'.
\end{equation}
where $x_i$ is the \emph{i}-th coordinate of the averaging point.

{If the local homogeneity hypothesis holds, $D_i$ can be shown to scale with $ D_{m,i}$, the size of the averaging domain in the \emph{i} direction  (see Appendix B). The scaling will depend also on the shape of the weighting function. }
%
{Moreover,  if the assumption of local homogeneity holds, $\mathbf{T}^\gamma$ contains all the dependence on the coarse-graining width of the stress tensor while the kinetic stress tensor $\mathbf{T}^k$ defined by means of the present fluctuation decomposition appears to be independent on averaging domain size.}

The translational kinetic energy balance is only slightly more complicated than the one obtained by Babic. Once more, most of the new terms are negligible under the local homogeneity assumption.

The total translational kinetic energy is found to be the sum of three contributions, $\rho E_T+\rho E^\gamma_T+\rho \varepsilon_T$,
 which are respectively the translational kinetic energy of the mean flow, 
the translational kinetic energy related to the velocity gradients, and the truly fluctuational translational kinetic energy.
These three variables are defined as
\begin{equation}
\rho E_T= \rho \frac{1}{2} \vec v \cdot \vec v,
\end{equation} 
\begin{equation}
\rho E^\gamma_T= \frac{1}{2} \rho (\vec D \cdot \nabla  \vec{\bar{v}})\cdot(\vec D \cdot \nabla \vec{\bar{v}}),
\end{equation}
\begin{equation}
\rho \varepsilon_T= \frac{1}{2} \int_{-\infty}^{\infty}\sum_p w_p m_p  \vec{\tilde{v}}_p   \cdot \vec{\tilde{v}}_p dt'.\label{eqn:epsilonT}
\end{equation} 
The translational kinetic energy balance equation is derived as:
\begin{eqnarray}\nonumber
\frac{\partial }{\partial t}(\rho E_T+\rho E^\gamma_T+\rho \varepsilon_T)   + \nabla \cdot(\rho E_T+\rho E^\gamma_T+\rho \varepsilon_T)  \vec{\bar{v}}=\\\nonumber
\nabla \cdot (\vec{q^k}+\vec{q^{c}})\\\nonumber+\nabla \cdot((\mathbf T^k +\mathbf T^c+\mathbf T^\gamma)\cdot \vec{\bar{v}}) \\\label{fluct}
+\rho \vec g \cdot \vec v-\gamma^k-\gamma^\gamma ,\;\;
\end{eqnarray}
where $\vec{q^k}$ and  $ \vec{q^c}$ are Babic's energy fluxes (see Appendix A). The last two terms correspond to
 the rate of conversion of kinetic energy to other forms of energy (dissipation due to friction, collisions, 
 plasticity, but also to storage and restitution of elastic energy). In the rigid limit they represent dissipation through
 friction and inelastic collisions and are therefore positive.
 The first one contains the contribution of fluctuations:
 \begin{equation}
\gamma^k= -\int_{-\infty}^{\infty}\sum_p\sum_{q>p}    \vec f_{pq} \cdot (\vec{\tilde{v_p}}- \vec{\tilde{v_q}}) w^S_{pq}dt',
\end{equation}
and the second one the rate of conversion of mechanical energy due to affine deformations:  
  \begin{equation}\label{gg}
\gamma^\gamma=-\int_{-\infty}^{\infty}\sum_p\sum_{q>p}   
 \vec f_{pq} \cdot ((\vec{x_p}-\vec{x_q})\cdot \nabla \vec{\bar{v}}) w^S_{pq}dt',
\end{equation}
where $w_{pq}^S$ is the mean of $w_p$ and $w_q$.
Both terms are independent of the size of the averaging domain.
We can see here again the benefit of our approach:  On one hand the domain size-dependent terms are clearly 
 identified in the equation ($\rho E^\gamma_T$ and $\mathbf T^\gamma$); on the other hand,
the contribution of affine deformations is separated from the contribution of the fluctuations. 
 
In fact, by superficially interpreting Babic's approach, one could conclude 
that all the energy is dissipated by means of fluctuations. This is incorrect, since affine deformations
do also dissipate energy; the reason for this misunderstanding is that affine
 deformations are considered as fluctuations in Babic's method. 

We can remove 
the mean flow kinetic energy terms by manipulating the 
momentum equation in a classical way (for example, see Ref.~\cite{woods75}).
The result is an equation expressing the balance for the 
fluctuating energy related to fluctuations and to affine deformations:
\begin{eqnarray}\nonumber
\frac{\partial }{\partial t}(\rho E^\gamma_T+\rho \varepsilon_T)     + \nabla \cdot (\rho E^\gamma_T+\rho \varepsilon_T)  \vec{\bar{v}}=\\\label{fluct2}
(\mathbf T^c+ \mathbf T^\gamma +\mathbf T^k )^\dagger : \nabla  \vec{\bar{v}}+\nabla \cdot ({\vec{q^k}+\vec{q^c}} ) -\gamma^k-\gamma^\gamma,
\end{eqnarray}
where $\dagger$ stands for transposition.
Let us do 
some more manipulation.
By looking at the term accounting for the rate of conversion of mechanical energy due to affine deformations, 
manipulation of Eq.~(\ref{gg}), making use of $\vec{l}_{pq}=\vec{x}_{q}-\vec{x}_{p}$, 
allows us to rewrite $\gamma^\gamma$ in the following form:
\begin{equation}\label{ggg}
\gamma^\gamma= \mathbf T^{c*\dagger} : \nabla \vec{\bar{v}},
\end{equation}
where
\begin{equation}
\mathbf{T}^{c*} =  \int_{-\infty}^{\infty}\sum_p\sum_{q>p}  w^S_{pq}  \vec{l}_{pq} \vec f_{pq} dt' , 
\end{equation}
 is a new contact stress tensor, differing from the $\mathbf{T}^{c}$ by the  weight used. Indeed, as mentioned above, $w^S_{pq}$ is the simple mean of $w_{p}$ and $w_{q}$, whereas $w^F_{pq}$ is related to the fraction of the branch vector $ \vec{l}_{pq}$ lying within the averaging region.
\section{One directional, steady, fully developed flow}
In this section we will apply our method to a simplified flow situation in order to gather some information on the effect 
of the new terms on force balance and on energy cascade.

Let us consider a one directional, steady, fully developed flow [defined by $\vec v= (v_x, 0, 0)$, $\partial_x=0$, $\partial_t=0$].
In such a case, the continuity equation is identically zero. The momentum equation becomes:
\begin{equation}\label{mom1D}
\nabla \cdot (\mathbf{T}^c+\mathbf{T}^\gamma + \mathbf{T}^k)  +\rho \vec g=0,
\end{equation}
and the fluctuating energy equation
\begin{equation}\label{fluc1D}
(\mathbf T^c+ \mathbf T^\gamma +\mathbf T^k )^\dagger : \nabla  \vec{\bar{v}}+\nabla \cdot ({\vec{q^k}+\vec{q^c}} ) -\gamma^k-\gamma^\gamma{=0.}
\end{equation}
According to the definition of $\mathbf T^\gamma$, it is easy to see that its only nonzero component is:
\begin{equation}\label{Tgxx}
\mathbf T^\gamma_{xx}=-\rho (D_y\partial_y v_x+D_z \partial_z v_x)^2,
\end{equation}
and it is therefore straightforward to see that 
\begin{equation}
 (\mathbf T^\gamma )^\dagger : \nabla  \vec{\bar{v}}=0,
\end{equation}
that is, the contribution of affine deformations to the stress tensor does not perform any work. 
Moreover, it is obvious that  $\mathbf T^\gamma$  directly affects only the
$x$ component of the  force balance. 
Given that the flow is fully developed,  we can safely remove it out from the balances, obtaining:
\begin{equation}\label{mom1D2}
\nabla \cdot (\mathbf{T}^c+ \mathbf{T}^k)  +\rho \vec g=0,
\end{equation}
\begin{equation}\label{fluc1D2}
(\mathbf T^c+ \mathbf T^k )^\dagger : \nabla  \vec{\bar{v}}+\nabla \cdot ({\vec{q^k}+\vec{q^c}} ) -\gamma^k-\gamma^\gamma{=0.}
\end{equation}
According to Eq.~(\ref{ggg}):
\begin{equation}
(\mathbf T^c)^\dagger : \nabla  \vec{\bar{v}} - \gamma^\gamma= (\mathbf T^c- \mathbf T^{c*}): \nabla  \vec{\bar{v}}.
\end{equation}
It is clear that $\mathbf T^c$ and $\mathbf T^{c*}$ are not in principle the same, particularly when the averaging domain is 
smaller than a particle diameter. However, if the averaging domain spans more than one particle diameter, we can assume that 
$\mathbf T^c\approx\mathbf T^{c*}$. 
Equation (\ref{fluc1D2}) becomes therefore:
\begin{equation}\label{fluc1D3b}
\mathbf T^k : \nabla  \vec{\bar{v}}+\nabla \cdot ({\vec{q^k}+\vec{q^c}} ) -\gamma^k{=0.}
\end{equation}
This means that the work provided by the contact part of the stress tensor is converted \emph{locally} by affine deformations, 
while only the kinetic part of the stress tensor enters into a nonlocal energy balance.
This result has important consequences, since it shows that separate paths may exist for the dissipation of the work of kinetic and contact stresses. However, it does not cast doubt on the validity of hydrodynamic theories~\cite{Bocquet_PRL_2002,JenkinsBerzi2010,artoni11,Tan_EPJE_2012} since our results do not deal with constitutive relations but suggest that a revision of the balances used for such theories could lead to a simplified description in some cases. 
{Concerning the size of the averaging domain, it may seem strange to talk about domains smaller than a  particle diameter, because 
it may seem obvious that such domains are inappropriate:
Few particles are indeed contained in a snapshot of such a domain. However, we are always dealing with space-time averages, so even if the averaging domain is small, in many cases (e.g., slowly evolving or steady-state processes) the domain indeed contains a lot of particles since many of them traveled through it during their history without affecting local homogeneity. In other cases, periodic boundary conditions may help too, allowing use of small domains along some directions.}
That is one of the reasons why in the next section we will employ the averaging method to some DEM simulation results 
pertaining to this class of simplified flows.
\section{DEM simulations}
\subsection{Method \& parameters}
\begin{figure}[t!]
\begin{center}
\includegraphics[width=0.74\columnwidth]{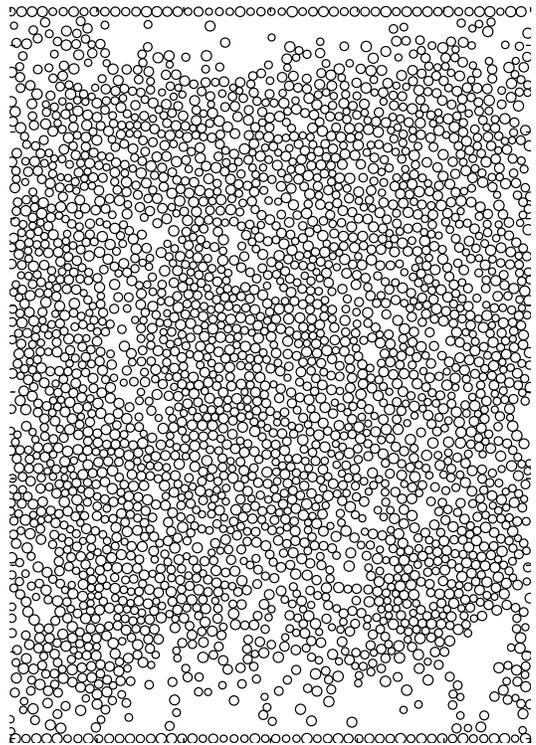}
\end{center}
\caption{Snapshot of the simulated system. A 2D granular material is sheared between two bumpy walls. The lower wall is fixed, the upper one moves horizontally at a constant velocity  $V_{top}=100$ and is submitted to a constant vertical stress $P=1$.}
\label{fig_sketch}
\end{figure} 
\begin{figure*}[ht!]
\begin{center}
\begin{tabular}{rcrc}
(a)&&(b)&\\
&\includegraphics[width=0.74\columnwidth]{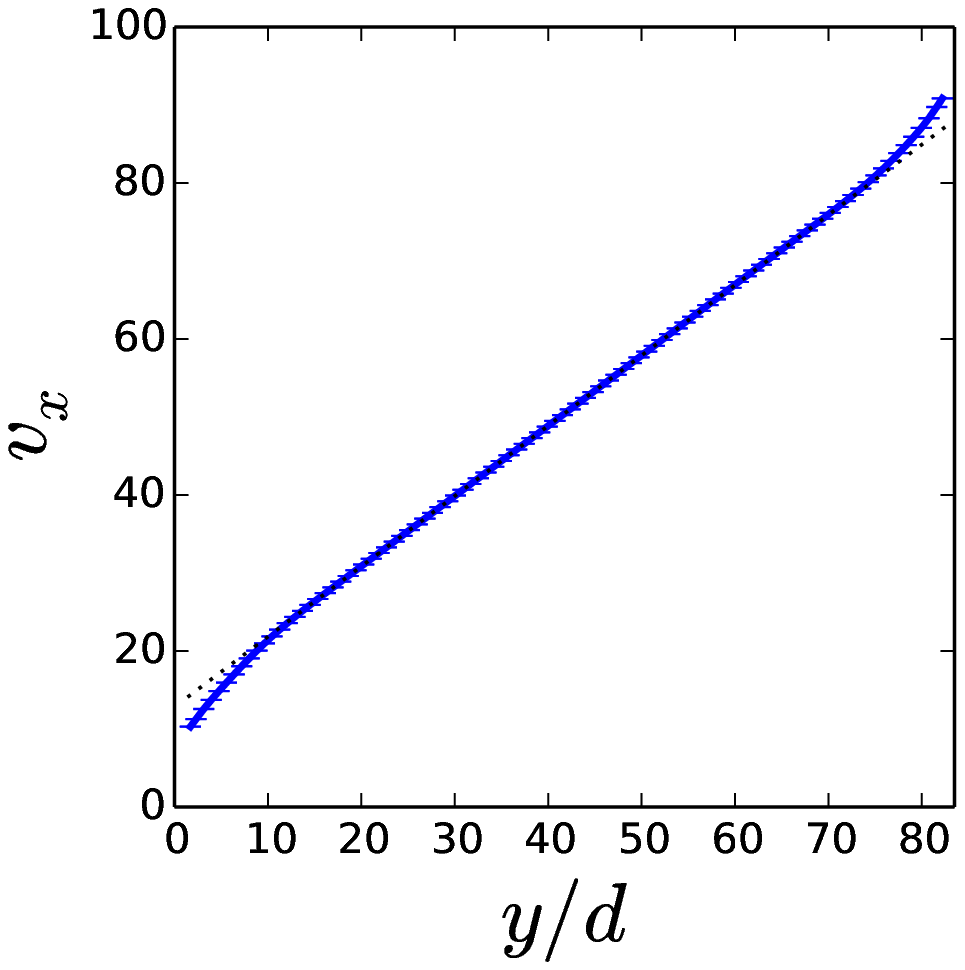}&&
\includegraphics[width=0.74\columnwidth]{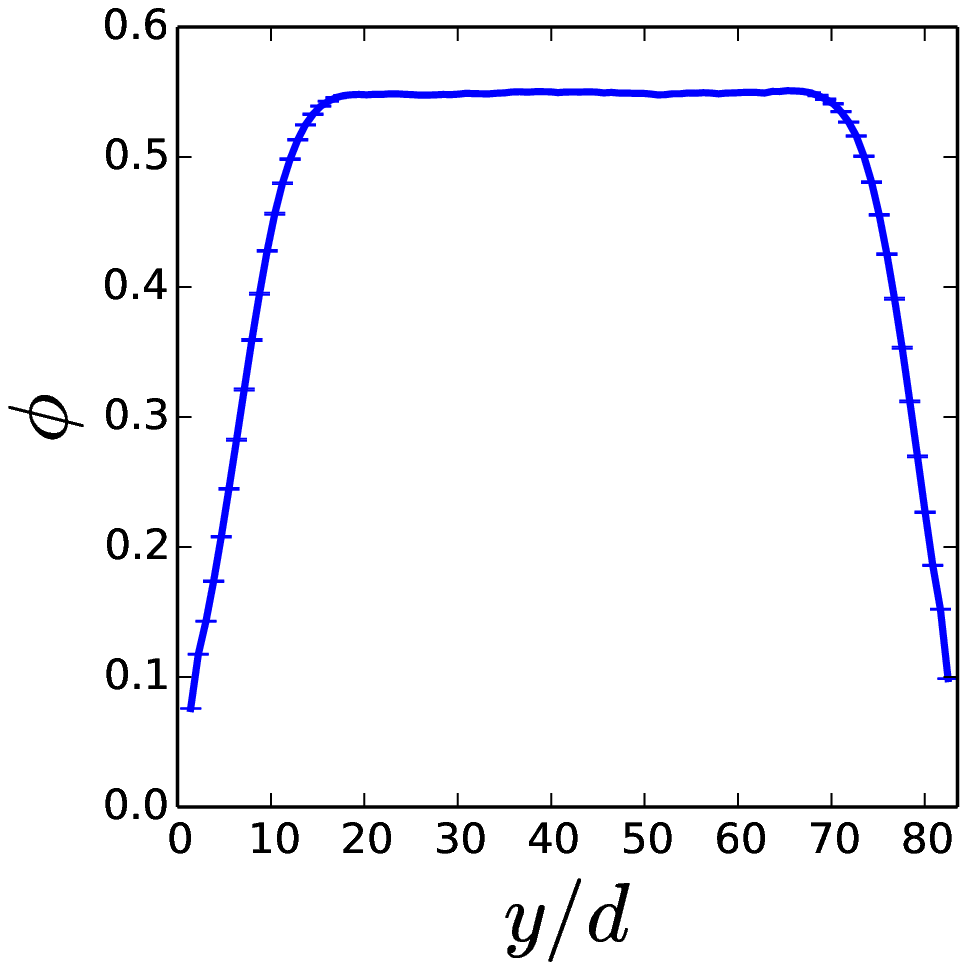} \\
\end{tabular}\end{center}
\caption{(Color online) (a) $x$ velocity and (b) solid fraction profiles along the $y$ direction  ($V_{top}=100$). The system appears to be in a simple shear condition  (linear velocity profile, constant solid fraction) in the center of the cell. The size of the averaging window is $2d$.}
\label{fig1}
\end{figure*}
\begin{figure*}[!t]
\begin{tabular}{rcrc}
(a)&&(b)&\\
&\includegraphics[height=0.74\columnwidth]{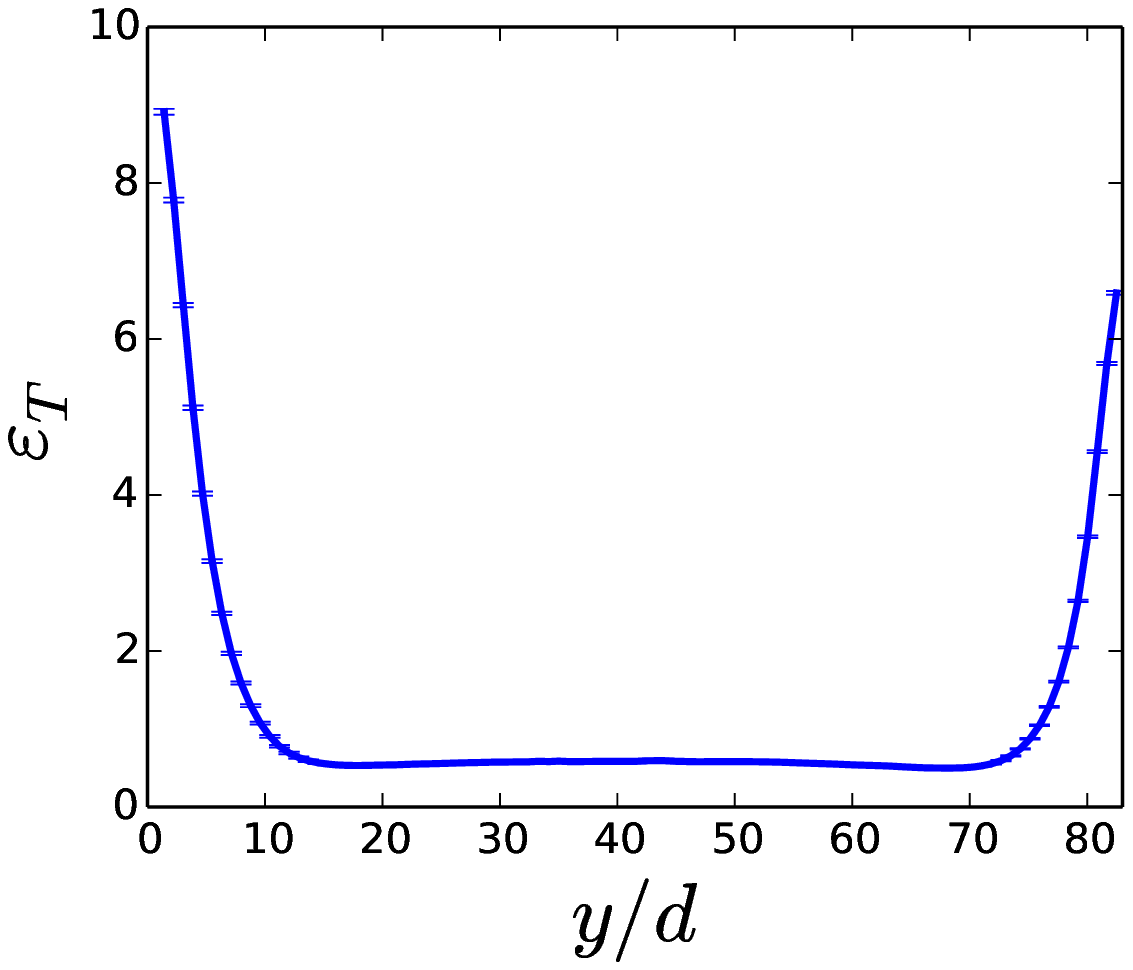}&&
\includegraphics[height=0.74\columnwidth]{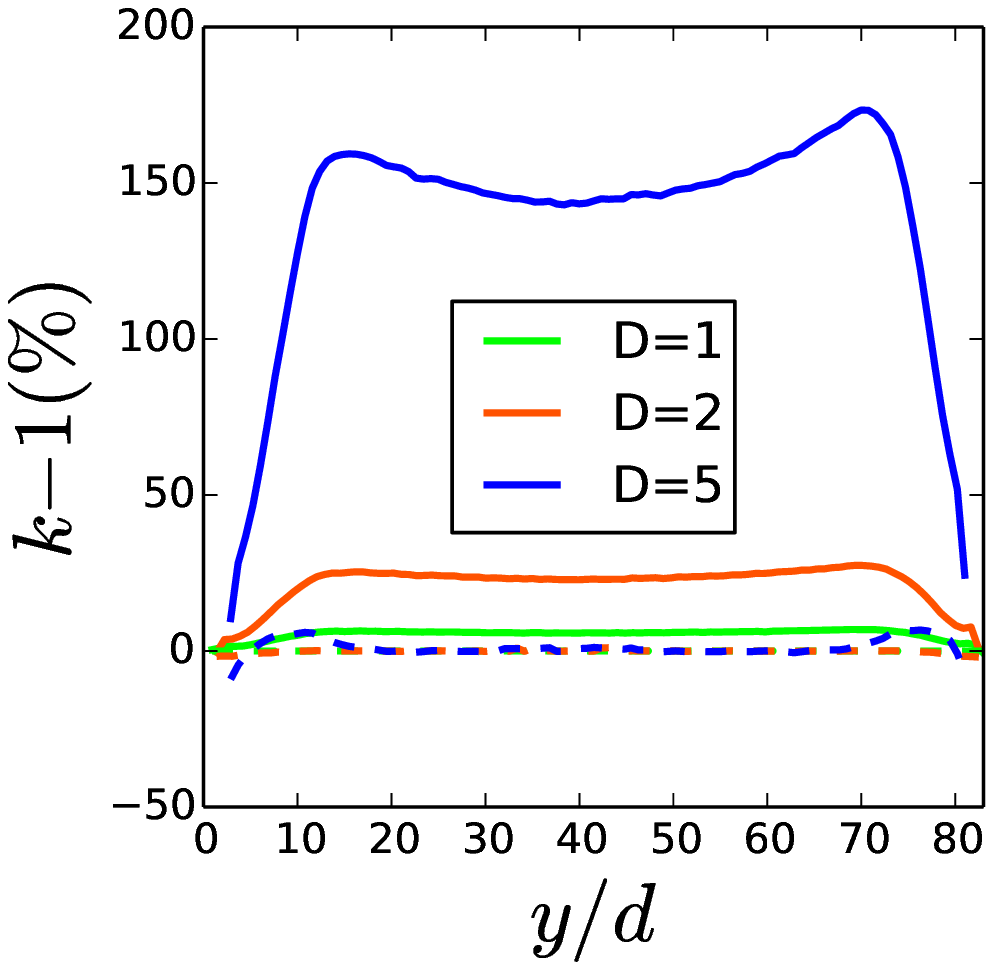} \\
\end{tabular}
\caption{ (Color online) (a) Fluctuating energy calculated by the present approach along the $y$ direction ($V_{top}=100$). The size of the averaging window is $2d$. 
(b) Relative difference (in percentage) between the fluctuating energy profiles obtained in several ways [Babic's ({solid} lines) or the present method ({dashed} lines), where {the color represents the size} of the averaging window] and that obtained with the present method and with an averaging window of $1d$.}
\label{fig2}
\end{figure*}

We use our own 2D implementation~\cite{Richard_PRL_2008,Richard2012} of the classical discrete element method where Newton's equations of motion for a system of $N$ ``soft'' disks are integrated.  
Such a technique, which is able to reproduce successfully the experimental results in many configurations (e.g., gravity-driven flows~\cite{Silbert2001,Richard_PRL_2008,Kumaran_PoF_2012,Kumaran_PoF_2013}, sheared systems~\cite{Rycroft2009,Vescovi_PoF_2014}, granular materials close to jamming~\cite{Majmudar2007}, silos~\cite{Hirshfeld2001}, and  rotating drums~\cite{Taberlet2006,Chand_PhysicaA_2012}), requires giving an explicit expression for the inter-particle forces. 
The discrete element method is classical and well known and can found in the aforementioned references. Therefore, we just present here the forces used in this work.\\
 
For the normal force between two overlapping disks we use a standard linear spring-dashpot interaction model~\cite{luding2008}, $F_n=k_n\delta_n-\gamma_n v_n$,
 where $\delta_n$ is the normal overlap, $k_n$ is the spring constant, $\gamma_n$ the damping coefficient, and $v_n$ the normal relative velocity. 
 The damping models the dissipation characteristic of granular materials. Likewise,  the tangential force is modeled as a linear elastic and linear dissipative force in the tangential direction
$F_t=k_t\delta_t - \gamma_t v_t$,   
 where $k_t$ is the tangential spring constant, $\delta_t$ the tangential overlap, $\gamma_t$  the tangential damping, and $v_t$ the tangential velocity at the contact point. 
 The magnitude of $\delta_t$ is truncated as necessary to satisfy the Coulomb law, $\left|F_t\right|\leq \mu \left| F_n \right|$, where $\mu$ is the grain-grain friction coefficient.
The simulated system is two dimensional (Fig.~\ref{fig_sketch}). The granular material is a dense assembly of $N$ dissipative disks of average diameter $d$ and average mass $m$. A small polydispersity of $\pm 20\%$ 
is considered to prevent crystallization.\\

The granular material is submitted to a plane shear, without gravity, leading to a uniform  stress distribution. 
The material is sheared between two parallel rough walls, separated by a distance $H$. 
One of the walls is fixed, while the other moves at the prescribed velocity $V$. 
The flow and transverse directions are respectively called $x$ and $y$. 
Periodic boundary conditions are applied along the flow direction, and  the length of the simulation box, $L_x$, is set to $60$ grain sizes. 
The wall roughness is made of disks sharing the characteristics of the flowing grains (same polydispersity and mechanical properties, no rotation). 
Their centers are equally spaced by a distance equal to the largest disk diameter $1.2 d$.
Along the $y$ axis, the position $y=0$ corresponds to the center of the glued grains on the fixed wall.
The normal stress applied on the moving wall, $P_0$, is controlled. The vertical position of the moving wall is thus not fixed and, 
using the method described in Ref.~\cite{DaCruz2005}, the height of the system $H$ obeys $\dot H =({P_0} - {P_w})L_x/g_p$, where $g_p$
is a viscous damping parameter and $P_w$ is the normal stress
exerted by the grains on the moving wall. 

The following values of
the parameters are used: $k_n/P_0= 10^5$, $k_t =
2k_n/7, \gamma_t = 0$, and $\mu = 0.5$ and $g_p=100\sqrt{k_n\,m}$. The value of $\gamma_n$ is adjusted to
obtain a normal restitution coefficient $e_n = 0.5$~\cite{Schafer1996}. 
The equations of motion for the translational and rotational
degrees of freedom are integrated with  a velocity-Verlet scheme with a time step $10^{-4}\sqrt{m/P_0}$. 
The number of grains, $N$,  is adjusted so the size of the system in the $y$ direction, $H$, is roughly equal to $80d$.
The initial state of the system is a randomly diluted
hexagonal lattice with disorder both on grain positions and velocities. 
The attainment of a steady state is verified by observing  time-invariant total kinetic
energy and distance between the two walls.

\subsection{Averaging procedure}

For the simulation set up described above, we implemented an averaging procedure following the method developed in the earlier sections of the paper. 

Several snapshots of the state of the system (positions, velocities, contact forces) at given times were captured. For each time a space average was performed; given that the system was periodic, averages were computed at different $y$ values.
A Heaviside step function was chosen as the weighting function. The only parameter of the weighting function was therefore the averaging window size. {Some authors suggested \cite{zhu02,Weinhart_PoF_2013}
smoother weighting functions. Here we preferred a step
function for its simplicity and to avoid smoothing  the profiles too
much. Given that the system was simple
and periodic, we felt that a nonsmooth function was the best choice, so as not to lose information about layering,
localization, and so on.}

Given that the simulations displayed a stationary state, the time average was performed as a simple mean over the results obtained for each snapshot. Standard errors of the time averages were estimated by the blocking method~\cite{flyvbjerg89} in order to take into account time correlation of the data.

Note that, in order to calculate the full set of variables, two steps were necessary: a first step to calculate average values of solid fraction, velocity, and contact stresses and a second step to calculate terms containing velocity fluctuations. Two steps are needed since velocity fluctuations are defined with respect to the mean field.

\subsection{Results}

In the following we will apply the method to a simulation with a velocity of the top wall $V_{top}=100 d\;\sqrt{P/m}$; as stated above, the flow is at constant pressure, so the vertical position of the top wall evolved freely 
 to a stationary value, $y_{top}\approx83.9 d$.

It is convenient to characterize such a shear using the so-called inertial number $I$ that compares the typical time scale of microscopic rearrangements with
the typical time scale of macroscopic deformations: $I=\dot\gamma d / \sqrt{P/\rho_S}.$ Note that $I$ corresponds to 
the square root of the Savage number~\cite{savage84} and is also called the Coulomb number~\cite{ancey99}.
Our system corresponds to a rapid flow with a global value of the inertial number corresponding to $I > 1$; such a flow regime was chosen because kinetic and contact stresses have the same order of magnitude, allowing us to test the results of the previous sections. A snapshot  of the system at steady state is given in Fig.~\ref{fig_sketch}. All profiles shown in the following are calculated using an averaging window of $2d$ unless otherwise stated. 

Figure~\ref{fig1}(a) shows the velocity profile at steady state. As highlighted in the figure, the profile is clearly linear except from two zones ($\sim 10 d$ wide) near the bumpy walls, 
where the shear rate increases approaching the boundary. 

The solid fraction profile [Fig.~\ref{fig1}(b)] displays a central zone with a constant value, and a decreasing behavior approaching the walls. 
The state of the system, characterized by a lower compaction near the walls, can be appreciated also from Fig. \ref{fig_sketch}.  
In fact, all the variables suggest that the system is in a simple shear condition (uniform shear, stresses, no energy flux) in the center 
and deviates from this state only in two narrow bands  ($\sim 10 d$ wide) near the bumpy walls. 
Another feature of the system that can be appreciated in Fig.~\ref{fig1}(b) is a slight asymmetry of the profiles near the walls. As it could be seen also in Fig.~\ref{fig_sketch}, in the shear zone near the upper wall the material is denser and less agitated than in the shear zone near the bottom wall. This slight difference is related to the asymmetry of boundary conditions: while the bottom wall is fixed, the upper one is free to move vertically in order to impose a constant normal stress. This yields slightly asymmetric profiles for all the variables.

\begin{figure}[!t]
\begin{center}
\begin{tabular}{rc}
(a)&\\
(b)&\includegraphics[width=.7\columnwidth]{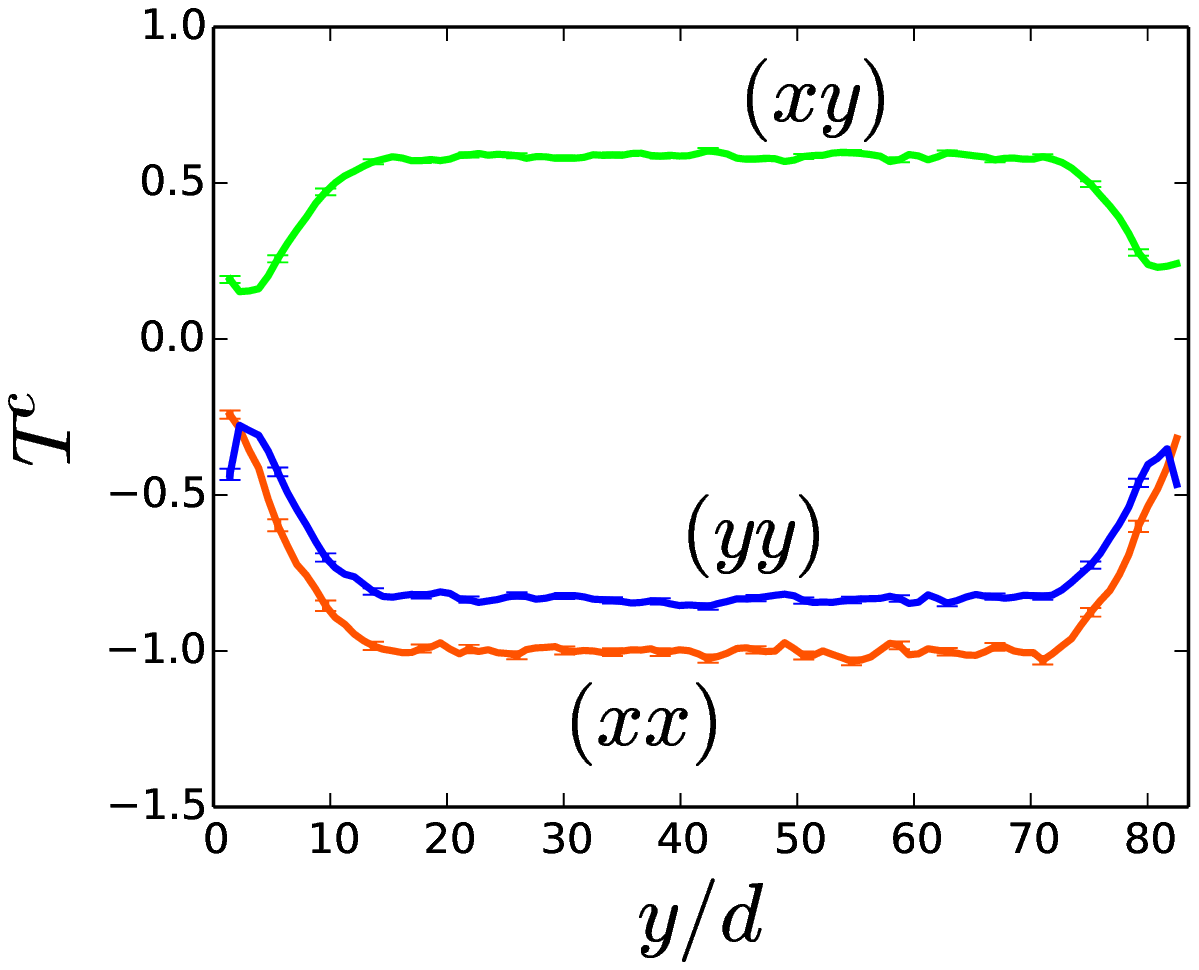}\\
(c)&\includegraphics[width=.7\columnwidth]{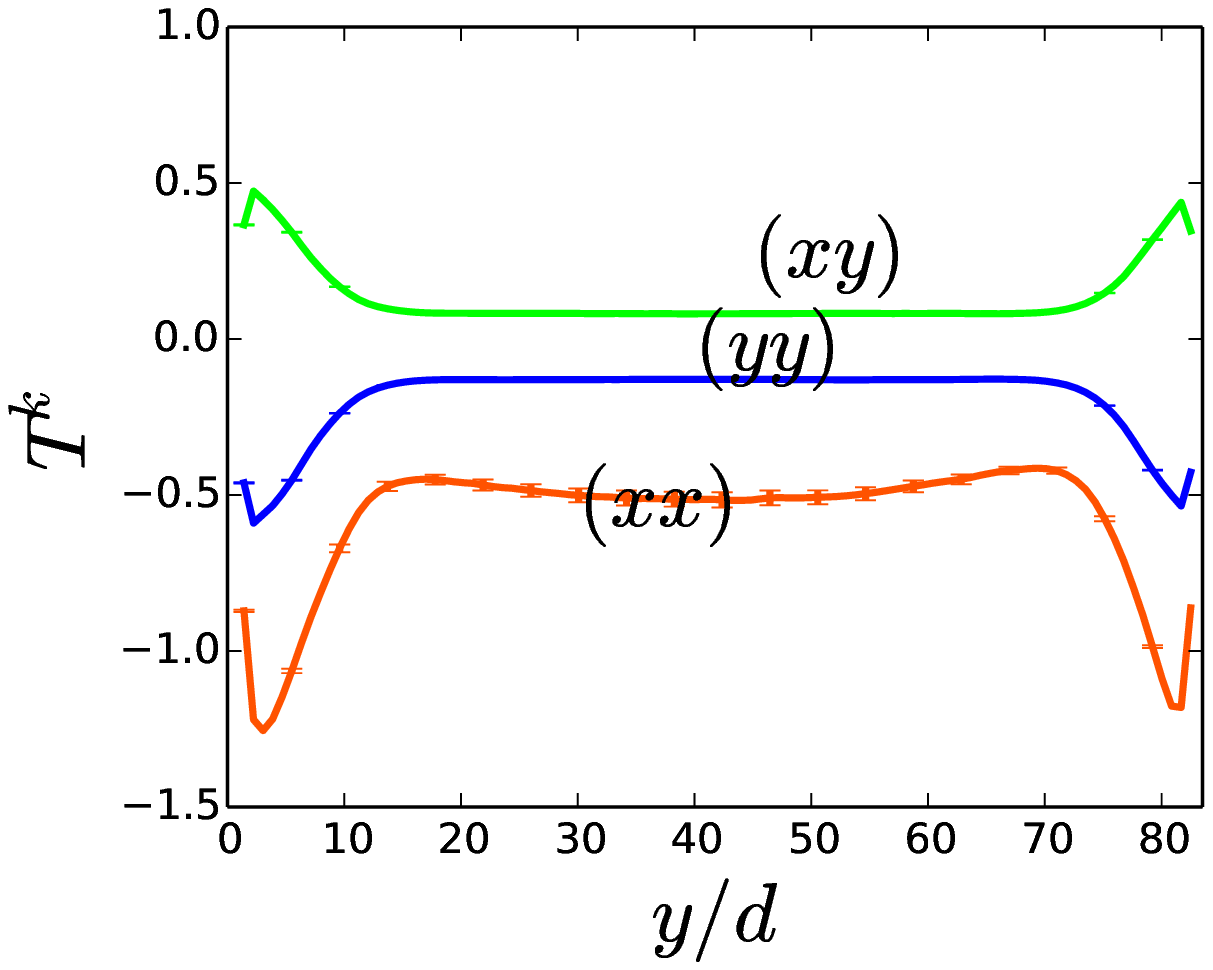}\\
&\includegraphics[width=.7\columnwidth]{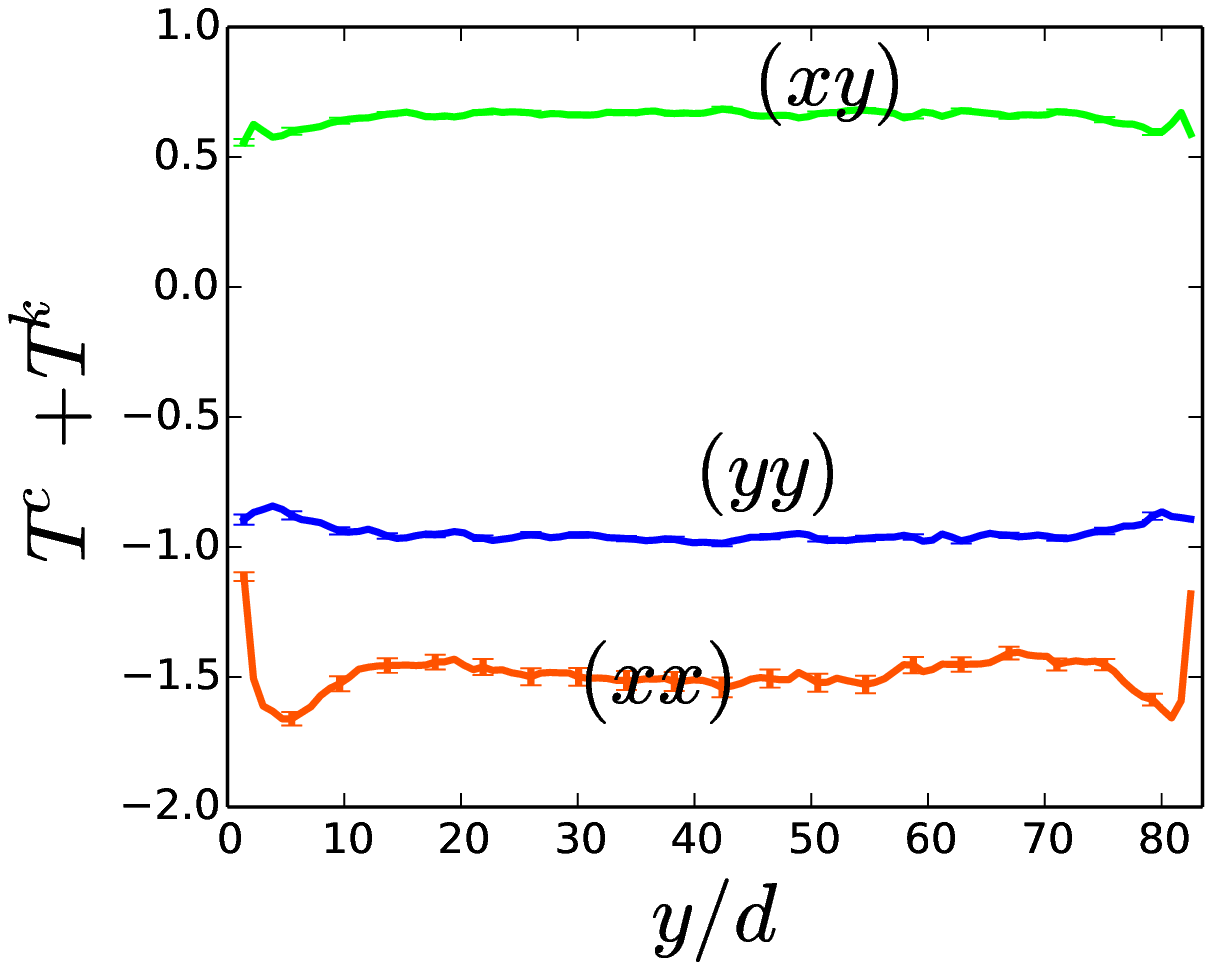}\\
\end{tabular}
\end{center}
\caption{(Color online) Components of the stress tensor [contact part (a), kinetic part (b), and total tensor (c)] along the $y$ direction. The upper wall velocity is $V_{top}=100$ and the size of the averaging window is $2d$.}\label{fig3}
\end{figure} 

\begin{figure*}
\begin{center}
\begin{tabular}{rcrc}
(a)&&(b)&\\
&\includegraphics[width=0.74\columnwidth]{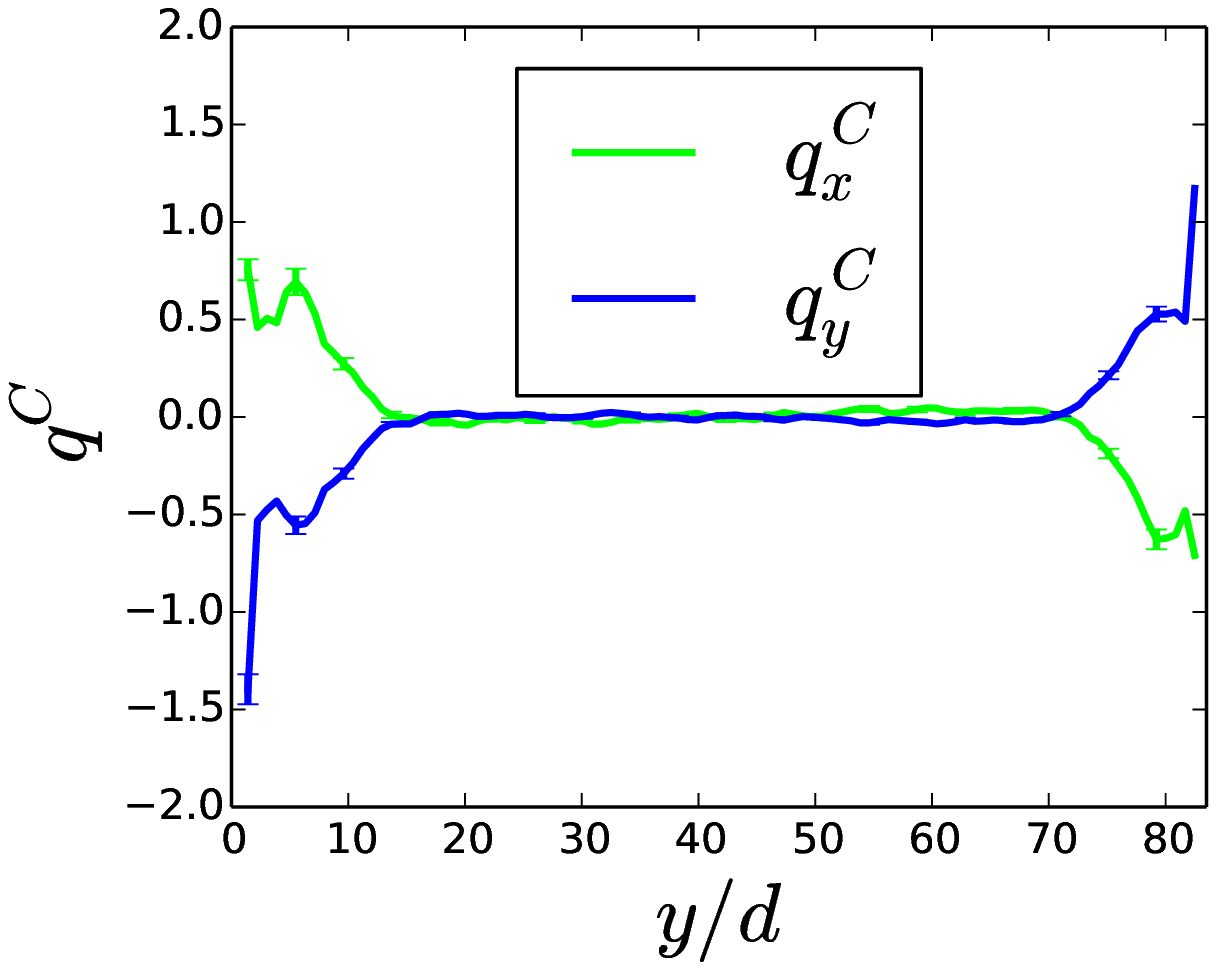}&&
\includegraphics[width=0.74\columnwidth]{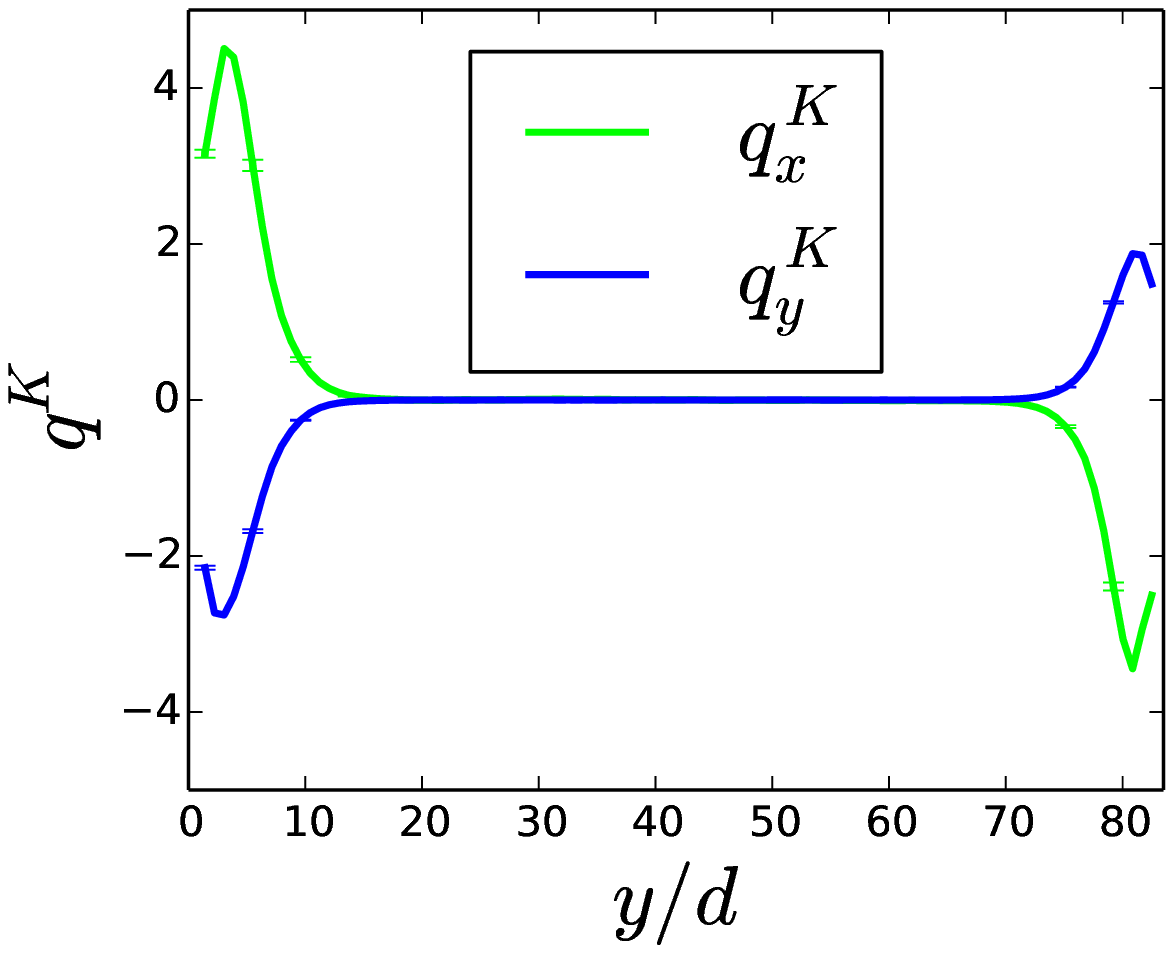}\\
\end{tabular}\end{center}
\caption{(Color online) Kinetic (a) and contact (b) fluxes of fluctuating energy along the $y$ direction  ($V_{top}=100$).  The size of the averaging window is $2d$.}
\label{figu}
\end{figure*} 
\begin{figure*}[t]
\begin{tabular}{rcrc}
(a)&&(b)&\\
&\includegraphics[width=0.74\columnwidth]{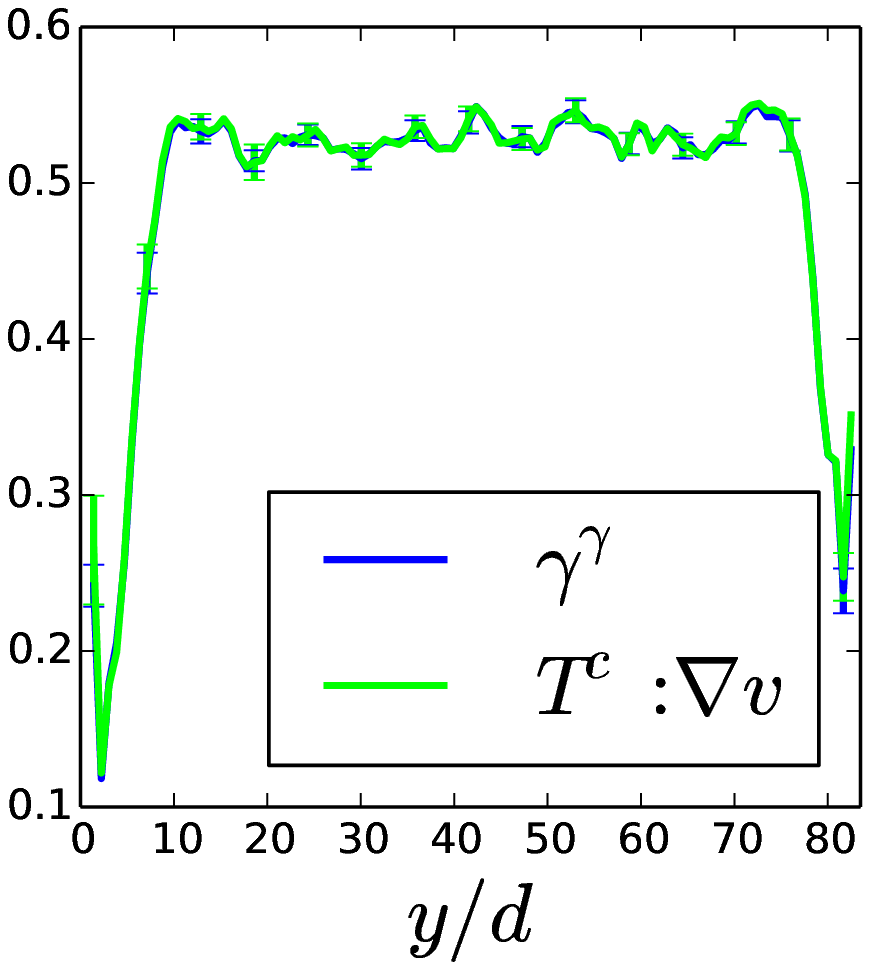}&&
\includegraphics[width=0.74\columnwidth]{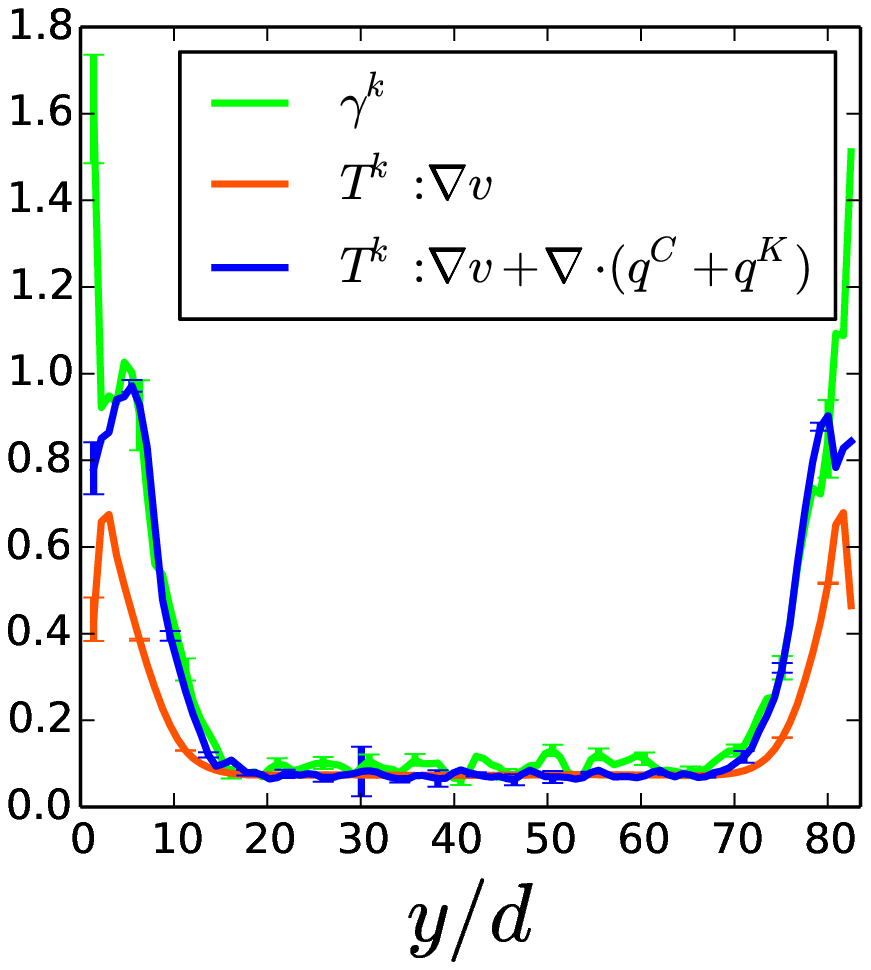} \\
\end{tabular}
\caption{(Color online) Stress power and dissipative terms: (a) contact stress tensor and (b) kinetic stress tensor, along the $y$ direction  ($V_{top}=100$).  The size of the averaging window is $2d$.}
\label{fig4}
\end{figure*} 
Figure \ref{fig2}(a) displays the fluctuating energy profile calculated with the new procedure developed in this work for an averaging window of $2d$.
To compare the two approaches (the present one and Babic's one) we report in Fig.~\ref{fig2}(b) the quantity
$k-1$, where $k$ is the ratio of the fluctuating energy profile calculated using one of the methods with different averaging sizes 
to a reference profile 
obtained with the present method and an averaging size equal to $D=1d$.
It clearly shows that the Babic's method leads to an estimate of the fluctuating energy which increases with the domain size. 
On the contrary, the present method leads to an estimate which does not depend on the size of the averaging window. However, if the latter size is too large it influences the averages close to the walls due to the inhomogeneity of the shear in such an averaging window. To illustrate this point, we have reported in Fig.~\ref{fig2}(b) results obtained with $D=5d$. Note also that, using the present method, the fluctuating energy profiles for $D=1d$ and $D=2d$ are almost equal for any position $y$.

\begin{figure*}[!ht]
\begin{tabular}{rcrc}
(a)&&(b)&\\
&\includegraphics[width=0.74\columnwidth]{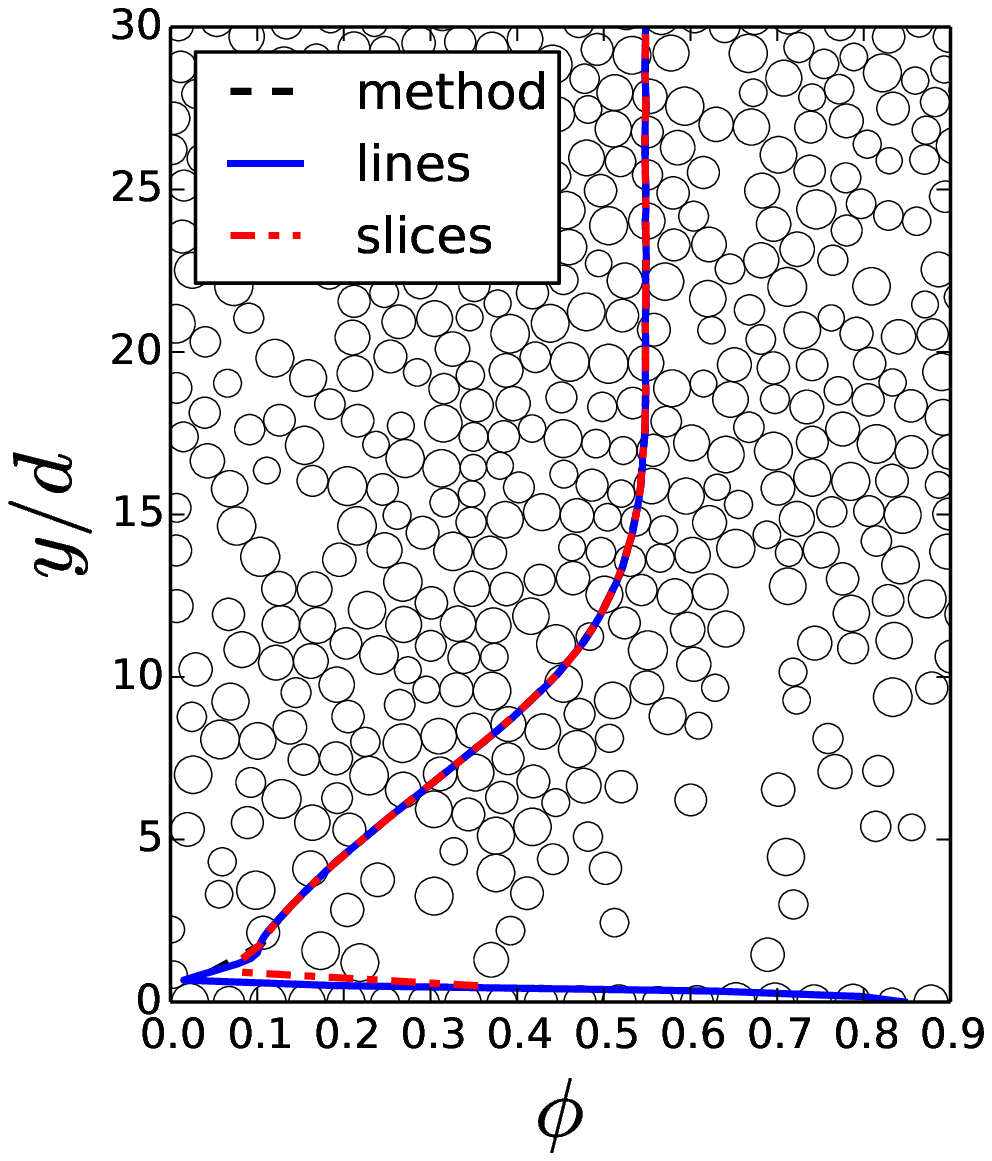}&&
\includegraphics[width=0.74\columnwidth]{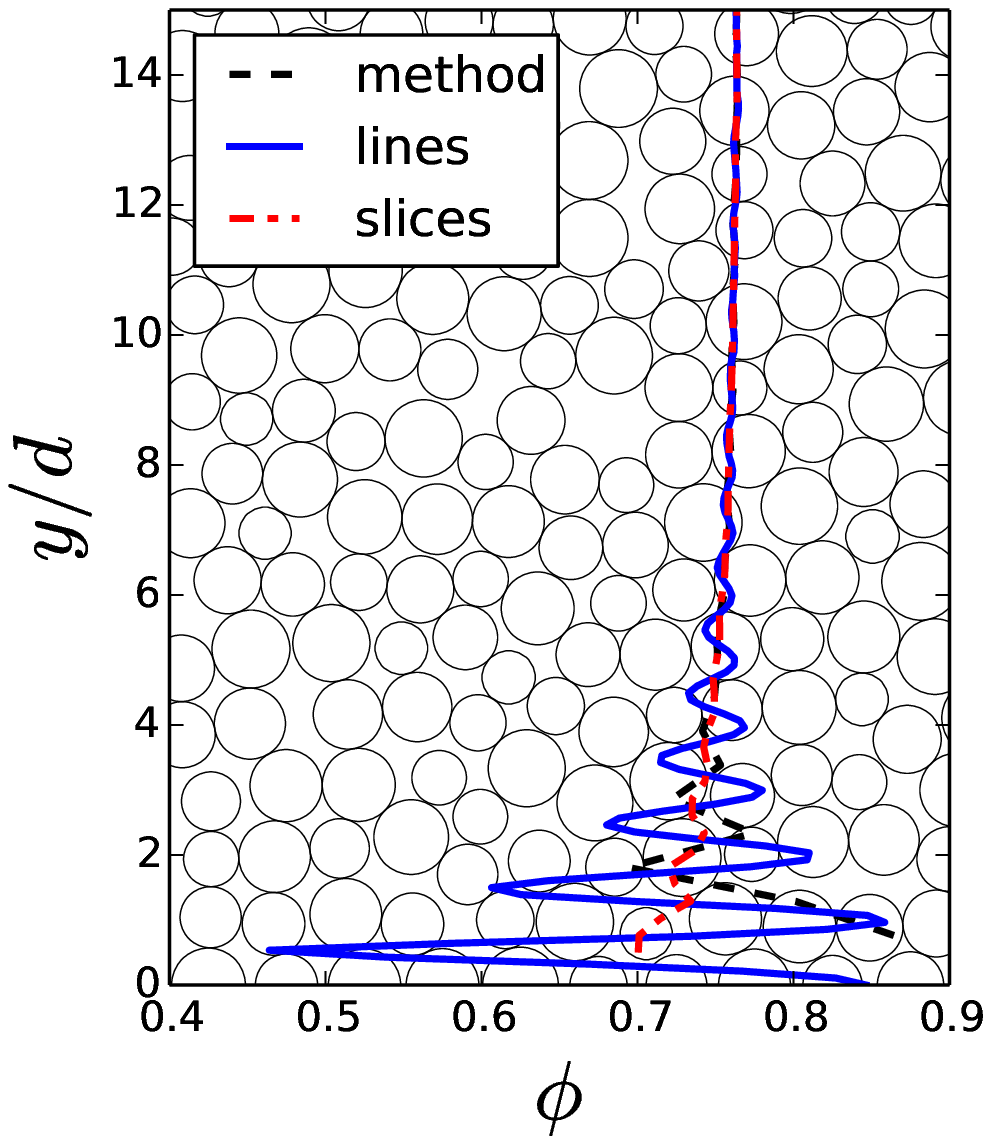}\\
\end{tabular}
\caption{(Color online) Layering close to a bumpy wall: solid fraction computed using the present method with an averaging window of $1d$, on lines and on slices ($1d$ wide) for (a) $V_{top}=100$, and (b) $V_{top}=10$.}
\label{fig5}
\end{figure*} 

Figure~\ref{fig3}(a)  displays the profiles of the components of the contact part of the stress tensor (here mainly due to collisions, since the flow is quite rapid). Again, a plateau is found at the center; when approaching the walls all the components decrease in absolute value.

At the same time  [Fig.~\ref{fig3}(b)],  the components of the kinetic part of the stress tensor increase when approaching the walls. 
This is a well-known direct consequence of the decreasing of solid fraction. In our simulation, near the walls, kinetic stresses are larger than collisional ones.

The sum of the two parts gives the total stress tensor, shown in Fig. \ref{fig3}(c). As predicted by the stress balance, the ($xy$) and ($yy$) components of the total stress tensor are uniform along $y$. 
{Concerning the third part of the stress tensor, $\mathbf T^\gamma$, it is clear that according to Eq.~(\ref{Tgxx}) its only nonzero component is the ($xx$) one. For the reference simulation it was verified that, apart for a small difference in a thin zone  ($\sim 2d$)  close to the walls (where slight deviations from local homogeneity appear due to the shape of the wall), $D_y \approx D_{m,y}/\sqrt{12}$, confirming the scaling given above. }

Figure \ref{figu} collects contact and kinetic flux profiles. All the components of these fluxes differ from zero only when approaching the wall; 
as regards the $y$ component of the fluxes, fluxes are negative when the fluctuating energy decreases with $y$ and positive otherwise. 
This agrees with the sign convention used for $\vec q$ in the derivation, and supports a  Fourier-like expression for the flux dependence on fluctuating energy gradient.
On the other hand, the horizontal components of the fluxes are not negligible near the walls. This contrasts with a linear transport theory since there is no gradient of fluctuating energy in the $x$ direction.  

Figure \ref{fig4} displays the estimates of the stress power related to contact and kinetic contributions and of the dissipative terms related to affine and non affine motions. 
Firstly, it can be seen that far from the walls, the algebraic relations hold
\begin{equation}
\mathbf T^{c\dagger} : \nabla  \vec{\bar{v}}=\gamma^\gamma, \mathbf T^{k\dagger} : \nabla  \vec{\bar{v}}=\gamma^k,
\end{equation}
therefore justifying the separate dissipation paths for kinetic and contact stress powers. 
Then, near the walls, the energy fluxes seem to influence only the kinetic part, therefore supporting the result obtained in the previous section: only kinetic terms enter in  a nonlocal energy balance, 
while contact stress power is dissipated locally.

\section{Discussion}
In the previous sections we developed an averaging procedure for granular materials which was based on a weighted average plus an original fluctuation decomposition.
Through numerical simulation results we proved the strength of the approach which gives consistent predictions for kinetic terms, 
and---more importantly---gives a new picture of the energy cascade in granular flows, at least for simple flows.
In this section we discuss the validity of such an approach.
It is clear that the assumption of local homogeneity holds only  when the solid fraction and the shear rate profiles do not display large gradients with respect to the averaging domain size.
For the simulations presented above, two zones were identified near the walls where such gradients exist. The independence of the fluctuating energy profile on the size of the averaging window  
and the respect of the stress balance are two proofs that the assumption is still acceptable. 

Figure \ref{fig5} allows us to discuss another important point: the physical meaning of the variables obtained by the method, with a particular regard to solid fraction. 
In the figure, the solid fraction profile was computed by two other methods:  estimating the fraction of  (1) a horizontal line  and (2) of a horizontal slice ($1 d $ wide) intersecting the particles.
These two methods have a clear geometrical meaning, the second being a moving average of the first.

For the rapid flow simulation which was taken as the reference [Fig. \ref{fig5}(a)], where no layering occurs, the three methods agree very well. 
Figure\ref{fig5}(b) displays the results of the three methods for a denser flow ($V_{top}=10$, $y_{top}\approx50$,$\Rightarrow I\approx 0.2$). Please note that the quite ordered bumpy wall was explicitly prepared to display layering. 
In this case the estimates on lines strongly fluctuate near the wall,~and the average on slices is a smoother profile which still fluctuates, while the estimate produced by the present averaging method seems to be 
intermediate between the two. Far from the wall results from the methods coincide.  Therefore, in the presence of layering, the solid fraction calculated by use of the present method may not have the direct geometrical
 meaning that its name implies. 
This problem may be reduced by choosing a smoother weighting function (in this paper we chose on purpose a very abrupt weighting function); 
we also suspect that in three dimensions this rather pathological issue may be less important.

Another issue caused by layering is that the strong spatial fluctuation of variables may imply that some of the assumptions of the local homogeneity hypothesis do not hold: 
The distribution of particles around the averaging point is not in principle homogeneous.
Fluctuations are evidently associated with strong local gradients of solid fraction and shear rate.
Therefore, when strong layering occurs, care has to be taken in applying the present method due to the approximations therein.

The inhomogeneous distribution of particles inside the averaging domain yields also a deviation of the virtual center of mass $\vec{X}$ from the center of the domain. This is not, in principle, a problem, because the averaging procedure is related to a volume and not to a point; average variables may be harmlessly referred to the virtual center of mass instead of the simple geometric center of the averaging domain. In other words the relation $\vec{X}=\vec{x}$ which seemed to be a consequence of the local homogeneity assumption can be taken as a definition of $\vec{x}$, the point where averages are supposed to be related.

In addition, results from simulations at $V_{top}=1$ and $V_{top}=10$ (which show quite strong layering), not shown here for the sake of brevity,  show that stress balance is respected and confirm locality of the dissipation of the contact stress power.  {However, 
if layering occurs, local homogeneity may be difficult to assume, and therefore the simplified balances obtained in this work have always to be critically analyzed when dealing with regions very close to the walls.}

\section{Conclusions} 
In this work, we propose a new derivation of continuum balance equations for granular materials which is inspired by the work of Babic~\cite{babic97}.
We introduce a new decomposition of particle scale velocities into mean and fluctuations, taking into account velocity gradients.
A very reasonable local homogeneity hypothesis allows us to simplify the balances, ensuring at the same time representativity and physical meaning of the continuum estimates obtained.

Two important results are obtained. 
First, the method solves the problem of dependence of some variables (e.g., fluctuating energy, kinetic stresses) on the averaging domain size which is intrinsic to previous methods~\cite{babic97} by clearly identifying scale-dependent terms.
Second, the development suggests a new physical picture for the energy cascade in granular flows: kinetic terms enter in a nonlocal energy balance, while contact terms (containing both collisions and long-lasting contacts) seem to be dissipated locally. 
Both results are verified against discrete element simulations of granular flows, with respect to which the method is shown at work.

This point is particularly important with regards to the 
hope of obtaining a statistical mechanics for granular flows whose predictions can be faithfully tested
in physical experiments and numerical simulations.
Note also that the question of the size of the averaging window close to a boundary~\cite{Weinhart_granularmatter_2012b} is problematic since it is bounded by the distance  between  the center of a grain and that boundary.  

A last note has to be added on the domain of application of the present method.
Discrete element methods are sometimes divided into smooth and nonsmooth ones~\cite{jean09}.
The present averaging method (as Babic's one) applies rigorously only to smooth methods, 
since the idea of smooth (differentiable) particle fields was implicit in the expression
of the equation of motion. The adaptation of such a method to non smooth dynamics is a very important
point which will be the subject of  future works. 

\section*{Acknowledgments}
We are indebted to M. Y. Louge for fruitful discussions and to S. McNamara for a critical reading of the manuscript. The numerical simulations  were carried out at the CCIPL (Centre de Calcul Intensif des Pays de la Loire) under the project MTEEGD.

\appendix
\section{Derivation of balance equations}
The derivation follows closely that of Babic, except for the choice of the fluctuation decomposition.
The evolution equation for a generic particle property $\psi_p$ is:
\begin{equation}\label{micro}
m_p\frac{d\psi_p}{dt}=\sum_q P_{pq} +m_p g_p.
\end{equation}
We multiply by $w_p$, sum on $p$, and integrate on time:
\begin{eqnarray}\nonumber
\int_{-\infty}^{\infty}\sum_p w_p m_p\frac{d\psi_p}{dt'} dt'=\\\label{first}
\int_{-\infty}^{\infty}\sum_p\sum_{q}  w_p  P_{pq}dt' +
\int_{-\infty}^{\infty}\sum_p  w_p m_p g_p dt'.
\end{eqnarray}
The left hand side term can be developed into~\cite{babic97}:
\begin{equation}\label{LHS}
\frac{\partial}{\partial t}\int_{-\infty}^{\infty}\sum_p w_p m_p\psi_p dt' +\nabla \cdot \int_{-\infty}^{\infty} \sum_p w_p m_p \psi_p \vec v_p dt',
\end{equation}
 
As regards the right hand side  of Eq.~(\ref{first}), Babic also showed that the interaction term could be developed as:
\begin{eqnarray}\nonumber
\int_{-\infty}^{\infty}\sum_p\sum_q  w_p  P_{pq}dt'=\\\nonumber
\int_{-\infty}^{\infty}\sum_p\sum_{q>p}  \left( P_{pq} + P_{qp}\right) w^S_{pq}dt'\\
+\nabla \cdot \frac{1}{2} \int_{-\infty}^{\infty}\sum_p\sum_{q>p} \vec{l}_{pq} \left( P_{pq} - P_{qp}\right)  w^F_{pq}dt'.
\end{eqnarray}
where $w^F_{pq}$ and $w^S_{pq}$ are two weighting functions respectively  defined as:
\begin{equation}
w^F_{pq}=  \int_{0}^{1}w(\vec{x}_p +s\vec{l}_{pq}-\vec{x},t'-t)ds,
\end{equation}
and
\begin{equation}
w^S_{pq}= \frac{w_p +w_q}{2}.
\end{equation} 
Now let us see the effect of the fluctuation decomposition on Eq.~(\ref{LHS}).
Decomposing the velocity field with the new decomposition leads to:
\begin{equation}
\vec v_p (\vec x_p)=\vec{\bar{v}} (\vec x) + (\vec x_p-\vec x) \cdot \nabla \vec{\bar{v}} (\vec x)+\vec{\tilde{v}}_p (\vec x_p).
\end{equation}
The divergence term on Eq.~(\ref{LHS}) can be written as:
\begin{eqnarray}\nonumber
\nabla \cdot \rho \bar{\vec{ v}} \bar \psi +\nabla \cdot \rho \overline{{\psi} \tilde{\vec{ v}}} \\ \label{div}+\nabla \cdot \left(\left(  \int_{-\infty}^{\infty} \sum_p w_p m_p \psi_p(\vec x_p-\vec x) dt'\right) \cdot \nabla \vec{\bar{v}} \right)
\end{eqnarray}
Decomposing also $\psi_p$ yields for the second term in Eq.~(\ref{div}):
\begin{eqnarray}\nonumber
\rho \overline{{\psi} \tilde{\vec{ v}}}=\left(\int_{-\infty}^{\infty} \sum_p w_p m_p\vec{\tilde{{v}}}_p(\vec x_p-\vec x)\right)
 \cdot \nabla \bar{\psi} (\vec x)dt '\\
 +\int_{-\infty}^{\infty} \sum_p w_p m_p\vec{\tilde{{v}}}_p\tilde{\psi_p}dt .
\end{eqnarray}
Given that \begin{equation}
\int_{-\infty}^{\infty} \sum_p w_p m_p  (\vec x_p-\vec x) dt'=\rho (\vec X - \vec x),
\end{equation}
where $\vec X$ is the average center of mass of the averaging domain, the integral in the third term of Eq.~(\ref{div}) becomes
\begin{eqnarray}\nonumber
 \int_{-\infty}^{\infty} \sum_p w_p m_p \psi_p(\vec x_p-\vec x) dt'=\bar{\psi} \rho  (\vec X - \vec x)\\\nonumber
+\int_{-\infty}^{\infty} \sum_p w_p m_p  ( (\vec x_p-\vec x) \cdot \nabla \bar{\psi})(\vec x_p-\vec x) dt'\\
+\int_{-\infty}^{\infty} \sum_p w_p m_p  \tilde{\psi_p}(\vec x_p-\vec x)dt'.
\end{eqnarray}

Following the homogeneity assumption, we neglect correlations of fluctuations and of 
particle positions, and we consider that the average
 center of mass coincides with the averaging point. This means that terms under
  the divergence operator in Eq.~(\ref{div}) simplify to 
\begin{eqnarray}\nonumber
\rho \bar{\vec{ v}} \bar \psi + \rho \overline{\tilde{\psi} \tilde{\vec{ v}}}+\\\label{div2}
\left(\int_{-\infty}^{\infty} \sum_p w_p m_p  ( (\vec x_p-\vec x) \cdot \nabla \bar{\psi})(\vec x_p-\vec x) dt'\right)\cdot \nabla\bar{\vec{v}}. 
\end{eqnarray}
Developing the $k$ component of the third term in Eq.~(\ref{div2})
[using the definition $\Delta x_{pi}= ( x_{pi}- x_i)$] yields:
\begin{eqnarray}\nonumber
\int_{-\infty}^{\infty} \sum_p w_p m_p  ( (\vec x_p-\vec x) \cdot \nabla \bar{\psi})((\vec x_p-\vec x)\cdot\nabla\bar{{v_k}} ) dt'=\\\label{kcomp}
\sum_i \sum_j \partial_{x_i}\bar{\psi} \partial_{x_j}{\bar{v_k}}\int_{-\infty}^{\infty} \sum_p w_p m_p  \Delta x_{pi}\Delta x_{pj}dt'.
\end{eqnarray}
Assumption of local homogeneity also implies decorrelation of the components of particle position with respect to the averaging point. This means that:
 \begin{equation}
\int_{-\infty}^{\infty} \sum_p w_p m_p  \Delta x_{pi}\Delta x_{pj}dt'=\rho D_i^2 \mbox{  if }i=j,\ 0 \mbox{ otherwise},
\end{equation}
where $D_i$ is related to the scale of the averaging volume in the $i$ direction. 
If $\vec D$ is the vector with components $D_i$, the third term of Eq.~(\ref{div2}) is
\begin{equation}\label{scdep}
\rho (\vec D \cdot \nabla \bar\psi) (\vec D \cdot \nabla \bar v_k).
\end{equation}
This term renders explicit the scale dependence which was otherwise implicit in Babic's method.
Equation~(\ref{LHS}) is finally reduced to:
\begin{equation}\label{LHS2}
\frac{\partial }{\partial t} \rho \bar \psi+ \nabla \cdot \rho \bar{\vec{ v}} \bar \psi +\nabla \cdot \rho \overline{\tilde{\psi} \tilde{\vec{ v}}}  +\nabla \cdot \left[\rho (\vec D \cdot \nabla \bar\psi)(\vec D \cdot \nabla \vec{\bar{v}})\right].
\end{equation}
As regards to the effect of the fluctuation decomposition on the right hand side of Eq.~(\ref{first}),  
 the result depends on the particular property $\psi_p$ considered. Thus in the following we will analyze 
 each property separately.
 
If mass conservation is considered ($\psi_p=1$, $P_{pq}=0$, $g_p=0$), the right-hand-side term of Eq.~(\ref{first}) is zero and it is straightforward to see that the 
classical continuity equation is obtained [Eq.~(\ref{conti}].

If force balances are considered ($\psi_p=\vec v_p$, $P_{pq}=\vec f_{pq}$, $g_p=\vec g$),  Eq.~(\ref{LHS}) becomes
\begin{equation}\label{lhsmom}
\frac{\partial }{\partial t} \rho \bar{\vec{ v}}+ \nabla \cdot \rho \bar{\vec{ v}} \bar{\vec{ v}} +\nabla \cdot \rho \overline{\tilde{\vec{v}} \tilde{\vec{ v}}}  +\nabla \cdot \left[\rho (\vec D \cdot \nabla \vec{\bar{v}})(\vec D \cdot \nabla \vec{\bar{v}})\right].
\end{equation}

The last two terms can be considered as related to two components of an effective stress tensor
 as defined by Eqs.~(\ref{tgamma}) and (\ref{tk}).
On the other hand, the right-hand side  of Eq.~(\ref{first}) is not affected by the nature of the fluctuation decomposition. Substitution 
of the  $P_{pq}=\vec f_{pq}$ and $g_p=\vec g$ yields to the standard momentum equation, Eq.~(\ref{momentum}), with the third 
component of the effective stress tensor given by Eq.~(\ref{tc}). 

Let us derive now the translational kinetic energy equation. In this case the reference particle property is
\begin{equation}
\psi_p = \frac{1}{2} \vec v_p \cdot \vec v_p.
\end{equation}
The average translational kinetic energy is
\begin{equation}
\rho \bar \psi =\frac{1}{2} \int_{-\infty}^{\infty}\sum_p  w_p m_p   \vec v_p \cdot \vec v_p dt',
\end{equation}
{and the fluctuation decomposition allows us to split this quantity  into six terms, three of which disappear due to the local homogeneity assumption, yielding:}
\begin{eqnarray}\nonumber
\frac{1}{2}\rho \bar{\vec{v}}   \cdot \bar{\vec{v}}  \\\nonumber +\frac{1}{2} \int_{-\infty}^{\infty}\sum_p w_p m_p
 ( (\vec x_p-\vec x) \cdot \nabla \vec{\bar{v}} )  \cdot ((\vec x_p-\vec x) \cdot \nabla \vec{\bar{v}}  ) dt' \\
+ \frac{1}{2} \int_{-\infty}^{\infty}\sum_p w_p m_p \vec{\tilde{v}_p}  \cdot\vec{\tilde{v}_p}  dt'
=\rho E_T+\rho E^\gamma_T+\rho \varepsilon_T,
\end{eqnarray}
where, with respect to Babic's development, $\rho \varepsilon_T$ is likely to be scale independent, and a new term appears, 
which is the translational kinetic energy related to affine deformation within the average volume $\rho E^\gamma_T$. 

The divergence term in Eq.~(\ref{LHS}) corresponds, for the kinetic energy, to:
\begin{equation}
\nabla \cdot \int_{-\infty}^{\infty} \sum_p w_p m_p \frac{1}{2}( \vec v_p\cdot  \vec v_p)  \vec v_p dt'.
\end{equation}
When developing this term, the fluctuation decomposition gives 27 terms, most of which can be deleted as in the
 previous development through (1) the postulate $\vec X=\vec x$, (2) the assumption of decorrelation among
  components of $\vec x _p - \vec x$, and (3) the assumption of decorrelation between components of
   $\vec x _p - \vec x$ and velocity fluctuations. 
It can be shown that the irreducible terms are 
\begin{equation}
(\rho E_T+\rho E^\gamma_T+\rho \varepsilon_T) \vec{\bar{v}} - \vec{q^k}- \mathbf{T}^k\cdot \vec{\bar{v}} - \mathbf{T}^\gamma\cdot \vec{\bar{v}}, 
\end{equation}
where 
\begin{equation}
\vec{q^k}= - \frac{1}{2}\int_{-\infty}^{\infty} \sum_p w_p m_p( \vec{{\tilde v}_p}\cdot  \vec{{\tilde v}_p}) \vec{{\tilde v}_p} dt'.
\end{equation}

Let us treat the terms coming from the right hand side term of Eq.~(\ref{first}). First, we must identify the particle scale interaction terms. 
Taking the product of $\vec{v_p}$ and Newton's equation, 
\begin{eqnarray}\nonumber
m_p \vec{v_p} \cdot \frac{d \vec{v_p}}{dt} = \vec{v_p} \cdot  (\sum_q \vec f_{pq} +m_p \vec g)\\
\Rightarrow m_p  \frac{d  \frac{1}{2} \vec{v_p}\cdot \vec{v_p}}{dt} = \vec{v_p} \cdot  (\sum_q \vec f_{pq} +m_p \vec g).
\end{eqnarray}
Therefore, the particle interaction term is 
\begin{equation}
 \sum_q   \vec f_{pq} \cdot \vec{v_p},
\end{equation}
and the particle source term is 
\begin{equation}
m_p   \vec g \cdot \vec{v_p}.
\end{equation}
Concerning the source term, local homogeneity implies:
\begin{equation}
\int_{-\infty}^{\infty}\sum_p  w_p m_p \vec g \cdot \vec v_p dt'=\rho \vec g\cdot \vec{\bar{v}}+(\rho(\vec X - \vec x)\cdot\nabla \vec{\bar{v}})\cdot \vec g=\rho \vec g\cdot \vec{\bar{v}}.
\end{equation}

Developing the interaction terms yields:
\begin{eqnarray}\nonumber
\int_{-\infty}^{\infty}\sum_p\sum_{q>p}  \left(  \vec f_{pq} \cdot \vec{v_p}+ \vec f_{qp} \cdot \vec{v_q}\right) w^S_{pq}dt'\\
+ \nabla \cdot \frac{1}{2} \int_{-\infty}^{\infty}\sum_p\sum_{q>p}  \left(  \vec f_{pq} \cdot \vec{v_p}- \vec f_{qp} \cdot \vec{v_q}\right)  \vec{l}_{pq} w^F_{pq}dt'.
\end{eqnarray}
Using the fluctuation decomposition, the relation $\vec f_{pq}=-\vec f_{qp}$, and the assumption of local homogeneity this becomes:
\begin{eqnarray}\nonumber
\int_{-\infty}^{\infty}\sum_p\sum_{q>p}    \vec f_{pq} \cdot (\vec{\tilde{v_p}}- \vec{\tilde{v_q}}) w^S_{pq}dt'\\\nonumber
+\int_{-\infty}^{\infty}\sum_p\sum_{q>p}    \vec f_{pq} \cdot ((\vec{x_p}-\vec{x_q})\cdot \nabla \bar v) w^S_{pq}dt'\\\nonumber
+ \nabla \cdot \frac{1}{2} \int_{-\infty}^{\infty}\sum_p\sum_{q>p}    \vec f_{pq} \cdot ( \vec{\tilde{v_p}}+\vec{\tilde{v_q}})  \vec{l}_{pq} w^F_{pq}dt'\\
+ \nabla \cdot \left[\left( \int_{-\infty}^{\infty}\sum_p\sum_{q>p}    \vec f_{pq}  \vec{l}_{pq} w^F_{pq}dt'\right)\cdot \vec{\bar v} \right],
\end{eqnarray}
which correspond respectively to:
\begin{equation}
-\gamma^k-\gamma^\gamma+\nabla \cdot \vec q^c +  \nabla \cdot \left(  \mathbf T^c \cdot \vec{\bar v}\right),
\end{equation}
where the first two terms are the conversion rates of translational kinetic energy into other forms of energy respectively due to velocity fluctuations ($-\gamma^k$) and to affine deformations ($-\gamma^\gamma$); $\vec q^c $ can be seen as a  translational kinetic energy flux related to contact forces.

The translational kinetic energy balance equation is therefore given by Eq.~(\ref{fluct}),
 which can be further simplified to Eq.~(\ref{fluct2}) when subtracting the mean flow energy balance.
{\section{Derivation of a scaling for $\vec{D}$}
In the following, a simplified scaling for the vector  $\vec{D}$ which enters in the definition of $\mathbf T^\gamma$ is derived using the local homogeneity hypothesis. Let us assume equal-sized spheres, and a step function on space and time as a weighting function. Let us consider variables computed at a point $(\vec x, t)$ in space and time. The average density is therefore:
\begin{equation}
\rho=m_p \int_{-\infty}^{\infty} \sum_p w_p   dt',
\end{equation}
 where $w_p=1/(VT)$ for particles residing in the averaging volume $V$ during the period $(t-T/2,t+T/2)$ and zero otherwise. If we call $N_p$ the time average of the number of particles lying in the averaging volume, it is clear that $\rho=m_pN_p/V$. 
Developing the definition of the vector $\vec D$, we find:
\begin{equation}
\rho D_i^2  =m_p \int_{-\infty}^{\infty} \sum_p w_p   (x_{pi}-x_i)^2 dt',
\end{equation}  
where $x_{pi}$ is the $i$ component of the particle position.
The local homogeneity hypothesis says that $\vec X=\vec x$ and that $x_{pi}$ is uniformly distributed around $x_i$. This yields for large $N_p T$ to:
\begin{eqnarray}\nonumber
\rho D_i^2  =\frac{m_p}{VT} \int_{-T/2}^{T/2} \sum_{p \in \Omega_{V,T}}   (x_{pi}-x_i)^2 dt'\\\approx \frac{m_p N_p T}{VT} \sigma_{x_{pi}-x_i}^2= \rho \frac{D_{m,i}^2}{12}.
\end{eqnarray}  
Therefore if the local homogeneity holds, $ D_i$ is likely to scale on the size of the averaging domain  in the $i$ direction. The proportionality will then depend on the choice of the weighting function, on polydispersity, on shape, and so on.
}


\end{document}